\documentclass[journal]{IEEEtran}

\usepackage{amssymb,amsthm}
\usepackage{amsmath}
\usepackage{csquotes}
\usepackage{caption}
\usepackage{subcaption}
\usepackage{hyperref,cleveref}
\usepackage{inconsolata}
\usepackage{listings}
\usepackage{mdframed}
\usepackage{url}
\usepackage[table]{xcolor}
\usepackage{xspace}
\usepackage{enumerate}
\usepackage{hyperref}
\usepackage{cleveref}
\usepackage{csquotes}
\usepackage{graphicx}
\usepackage{mdframed}
\usepackage{xcolor}
\usepackage{xspace}
\usepackage{inconsolata}
\usepackage{verbatim}
\usepackage{fancyvrb}
\usepackage{todonotes}
\usepackage{multirow}

\newcommand{\icol}[1]{
  \left(\begin{matrix}#1\end{matrix}\right)%
}

\usepackage{listings}

\makeatletter

\newcommand\PrologPredicateStyle{}
\newcommand\PrologVarStyle{}
\newcommand\PrologAnonymVarStyle{}
\newcommand\PrologAtomStyle{}
\newcommand\PrologOtherStyle{}
\newcommand\PrologCommentStyle{}

\newif\ifpredicate@prolog@
\newif\ifwithinparens@prolog@

\lst@SaveOutputDef{`_}\underscore@prolog

\newcount\currentchar@prolog

\newcommand\@testChar@prolog%
{%
  \ifnum\lst@mode=\lst@Pmode%
    \detectTypeAndHighlight@prolog%
  \else
    \ifwithinparens@prolog@%
      \detectTypeAndHighlight@prolog%
    \fi
  \fi
  \global\predicate@prolog@false%
}

\newcommand\detectTypeAndHighlight@prolog
{%
  \def\lst@thestyle{\PrologAtomStyle}%
  \ifpredicate@prolog@%
    \def\lst@thestyle{\PrologPredicateStyle}%
  \else
    \expandafter\splitfirstchar@prolog\expandafter{\the\lst@token}%
    \expandafter\ifx\@testChar@prolog\underscore@prolog%
      \ifnum\lst@length=1%
        \let\lst@thestyle\PrologAnonymVarStyle%
      \else
        \let\lst@thestyle\PrologVarStyle%
      \fi
    \else
      \currentchar@prolog=65
      \loop
        \expandafter\ifnum\expandafter`\@testChar@prolog=\currentchar@prolog%
          \let\lst@thestyle\PrologVarStyle%
          \let\iterate\relax
        \fi
        \advance \currentchar@prolog by 1
        \unless\ifnum\currentchar@prolog>90
      \repeat
    \fi
  \fi
}
\newcommand\splitfirstchar@prolog{}
\def\splitfirstchar@prolog#1{\@splitfirstchar@prolog#1\relax}
\newcommand\@splitfirstchar@prolog{}
\def\@splitfirstchar@prolog#1#2\relax{\def\@testChar@prolog{#1}}

\def\beginlstdelim#1#2%
{%
  \def\endlstdelim{\PrologOtherStyle #2\egroup}%
  {\PrologOtherStyle #1}%
  \global\predicate@prolog@false%
  \withinparens@prolog@true%
  \bgroup\aftergroup\endlstdelim%
}

\newcommand\lang@prolog{Prolog-pretty}
\expandafter\lst@NormedDef\expandafter\normlang@prolog%
  \expandafter{\lang@prolog}

\expandafter\expandafter\expandafter\lstdefinelanguage\expandafter%
{\lang@prolog}
{%
  language            = Prolog,
  keywords            = {},      
  showstringspaces    = false,
  alsoletter          = (,
  alsoother           = @$,
  moredelim           = **[is][\beginlstdelim{(}{)}]{(}{)},
  MoreSelectCharTable =
    \lst@DefSaveDef{`(}\opparen@prolog{\global\predicate@prolog@true\opparen@prolog},
}

\newcommand\@ddedToOutput@prolog\relax
\lst@AddToHook{Output}{\@ddedToOutput@prolog}

\lst@AddToHook{PreInit}
{%
  \ifx\lst@language\normlang@prolog%
    \let\@ddedToOutput@prolog\@testChar@prolog%
  \fi
}

\lst@AddToHook{DeInit}{\renewcommand\@ddedToOutput@prolog{}}

\makeatother
\definecolor{PrologPredicate}{RGB}{000,031,255}
\definecolor{PrologVar}      {RGB}{024,021,125}
\definecolor{PrologAnonymVar}{RGB}{000,127,000}
\definecolor{PrologAtom}     {RGB}{153,034,085}
\definecolor{PrologComment}  {RGB}{063,128,127}
\definecolor{PrologOther}    {RGB}{000,000,000}

\renewcommand\PrologPredicateStyle{\color{PrologPredicate}}
\renewcommand\PrologVarStyle{\color{PrologVar}}
\renewcommand\PrologAnonymVarStyle{\color{PrologAnonymVar}}
\renewcommand\PrologAtomStyle{\color{PrologAtom}}
\renewcommand\PrologCommentStyle{\itshape\color{PrologComment}}
\renewcommand\PrologOtherStyle{\color{PrologOther}}

\lstdefinestyle{Prolog-pygsty}
{
  language     = Prolog-pretty,
  upquote      = true,
  stringstyle  = \PrologAtomStyle,
  commentstyle = \PrologCommentStyle,
  literate     =
    {:-}{{\PrologOtherStyle :-}}2
    {,}{{\PrologOtherStyle ,}}1
    {.}{{\PrologOtherStyle .}}1
}

\makeatletter
\lst@Key{countblanklines}{true}[t]%
    {\lstKV@SetIf{#1}\lst@ifcountblanklines}

\lst@AddToHook{OnEmptyLine}{%
    \lst@ifnumberblanklines\else%
       \lst@ifcountblanklines\else%
         \advance\c@lstnumber-\@ne\relax%
       \fi%
    \fi}
\makeatother

\newcommand{\rev}[1]{\textcolor{black}{#1}\xspace}

\newcommand{\eg}{e.g.,\xspace}

\newcommand{\policy}[1]{\textsf{\small #1}\xspace}
\newcommand{\alwayslow}{\policy{always low}}
\newcommand{\alwaysmedium}{\policy{always medium}}
\newcommand{\naive}{\policy{naïve}}

\crefname{equation}{Eq.}{Eqs.}

\newcommand{\carbostat}{\textsc{Car\-bon\-stat}\xspace}

\begin{document}
\title{Carbon-aware Software Services}

\author{
    Stefano Forti, 
    Jacopo Soldani, and 
    Antonio Brogi
    \thanks{S. Forti, J. Soldani, and A. Brogi are with the Department of Computer Science, University of Pisa, Italy. E-mail: \texttt{\{name.surname\}@unipi.it}.}
}

\markboth{Submitted for publication}{} 

\maketitle

\begin{abstract}
The significant carbon footprint of the ICT sector calls for methodologies to contain carbon emissions of running software. This article proposes a novel framework for implementing, configuring  and assessing carbon-aware interactive software services.
First, we propose a methodology to implement carbon-aware services leveraging the Strategy design pattern to feature alternative service versions with different energy consumption. Then, we devise a bilevel optimisation scheme to configure which version to use at different times of the day, based on forecasts of carbon intensity and service requests, pursuing the two-fold goal of minimising carbon emissions and maintaining average output quality above a desired set-point. 
Last, an open-source prototype of such optimisation scheme is used to configure a software service implemented as per our methodology and assessed against traditional non-adaptive implementations of the same service. Results show the capability of our framework to control the average quality of output results of carbon-aware services and to reduce carbon emissions from 8\% to 50\%.
\end{abstract}


%
\IEEEpeerreviewmaketitle

\section{Introduction}
\label{sec:intro}\noindent
The Information and Communication Technologies (ICT) sector is producing around 2\% of the worldwide carbon emissions, i.e. as much as airplane traffic, with an energy consumption estimated at between 6\% and 9\% of the global demand and expected to grow up to 20\% by 2030~\cite{belkhirAssessing2018,carbonict2023}. Such exponential growth is due to the increase in the amounts of data produced, transmitted, and consumed worldwide\footnote{\url{https://www.statista.com/statistics/871513/worldwide-data-created/}} through interactive \rev{software} services (e.g., video streaming services, social networks, and Internet-of-Things-based applications).  This, along with the urge to reduce carbon emissions to ensure a future for our Planet by meeting the Paris agreement\footnote{\url{https://www.un.org/en/climatechange/paris-agreement}} objectives, poses non-trivial challenges. 

Up to the early 2010s, the empirical law known as \textit{Dennard scaling} ensured that the power consumption of an integrated circuit of a given area is constant independently from the number of transistors it contains~\cite{dennard1974}.
In parallel, Moore's law observed that the number of transistors in a chip would double approximately every 18 months~\cite{Moore98a}. Following these, big players in the Cloud industry upgrade their datacentres' hardware approximately every two years to double their computing power without increasing energy costs and reducing their carbon emissions per unit of computation. Much research has lately targeted the optimisation of renewable energy usage through suitable service orchestration and consolidation~\cite{GaglianeseGreenOrch2023} .

\rev{However, with Dennard's scaling and Moore's law slowing down~\cite{DBLP:journals/csur/MuralidharBB22}, further improvements towards a more sustainable future for ICT will not be consistently supported by hardware improvements but will require adopting suitable energy- and carbon-aware software engineering tools and methodologies that are currently lacking, i.e. treating ``\textit{energy and carbon as a first-class resource}"~\cite{treehouse2023} in our code. Similarly, Verdecchia et al.~\cite{DBLP:journals/software/VerdecchiaLEVE21} call for methodologies to write sustainable software by blending decision-making with the software execution context.}
%

This article 
proposes a framework to implement, configure and assess carbon-aware \rev{interactive (i.e. request-response)} \rev{software} services, made from the following: 
\begin{enumerate}
    \item[(\textit{i})] a \textit{methodology}, illustrated via a case study, to implement software services by relying on the \textit{Strategy} design  pattern~\cite{designpatternsgamma} to feature alternative implementations of the same \rev{functionality}, each associated with their own output \textit{quality scores} (e.g., error, QoE) and \textit{execution times},
    \item[(\textit{ii})] an \textit{optimisation schema} and its Python \textit{prototype}\footnote{Open-sourced and freely available at: \url{https://github.com/di-unipi-socc/carbonstat/tree/main}}, \carbostat, to suitably \textit{configure} such adaptive \rev{software} services by selecting which \rev{functionality} implementation to use so as to minimise carbon emissions and ensure target average output quality, and
    \item[(\textit{iii})] an \textit{experimental assessment} of different service configurations obtained via \carbostat for a \rev{software} service implemented as per the proposed methodology, and relying on real carbon intensity data and lifelike requests' patterns as contextual data for the experiments.
\end{enumerate}

\begin{figure}[!t]
    \centering
    \includegraphics[width=0.37\textwidth]{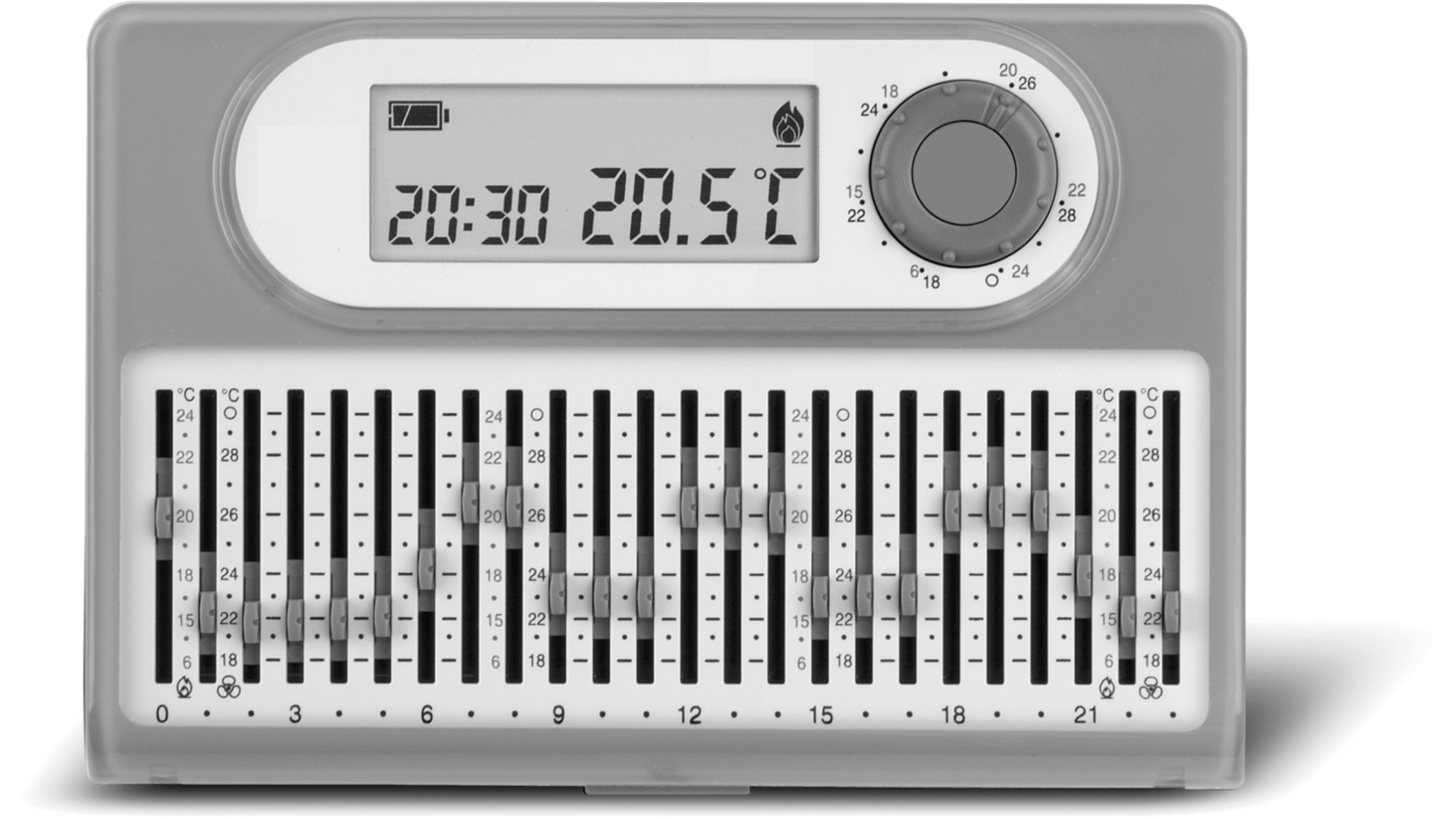}
    \caption{A thermostat with sliders.}
    \label{fig:thermostat}
\end{figure}

This work enables developers to implement different versions of the same request-response service, retaking the baseline idea of approximate computing~\cite{approximatecomputingMittal16b}, i.e., that shorter execution times usually result in lower energy consumption at the price of lower output quality scores. By considering this, the forecast service requests and the carbon intensity of the energy mix powering up deployment servers, \carbostat suitably selects the version to run at different times of the day to minimise the carbon emissions of a running service and to keep the average results' quality above a set-point. \rev{Notably, our methodology can also be used to reconfigure existing application services so as to accommodate carbon-awareness, aiming at reducing their environmental footprint. }

By analogy, consider thermostats that measure the temperature of a building and trigger its heating system to maintain such temperature near a desired set-point throughout the day. Some thermostats (Fig.~\ref{fig:thermostat}) allow users to move sliders to specify different set-points for different times of the day, depending on their thermal comfort preferences, presence in the building, and weather forecasts.
Similarly, based on the forecasts of the carbon intensity of the employed energy mix and the number of incoming service requests, through mixed integer linear programming (MILP), \carbostat decides which version (viz. strategy) to use of a certain service at different times of the day so to minimise carbon emissions due to service computation and keep the average result quality close to a set threshold. 

\carbostat considers the average error on the returned output as the quality metric to optimise. It is assessed against baseline four different baseline configurations by relying on forecast and real data traces from the carbon intensity of the United Kingdom\footnote{Data available at: \url{https://carbonintensity.org.uk}} from 2023.  
Results show how \carbostat -- fed with reliable forecasts -- can ensure consistent reduction on carbon emissions and average error rates within set thresholds under different request load conditions, whereas static naive policies might incur in unpredictable behaviour. Note that the employed experimental setup and code can also be used to assess and fine-tune services designed as per our methodology,   by varying contextual data.

The rest of this article is organised as follows. We illustrate, through a case study, our methodology to implement carbon-aware interactive \rev{software} services (Sect.~\ref{sec:design}). Then, we describe the MILP formulation of \carbostat to automatically configure the behaviour of such services, based on forecast contextual data and target quality of output results (Sect.~\ref{sec:carbomilp}). Subsequently, we assess the use case service, configured via \carbostat, by relying on forecast and real data for carbon intensity and on a set of lifelike request patterns (Sect.\ref{sec:experiments}). We conclude by discussing some related efforts (Sect.~\ref{sec:related}) and directions for future work (Sect.~\ref{sec:conclusions}).

\section{Carbon-aware Service Design}
\label{sec:design}\noindent

\begin{figure}
    \centering
    \includegraphics[width=0.5\textwidth]{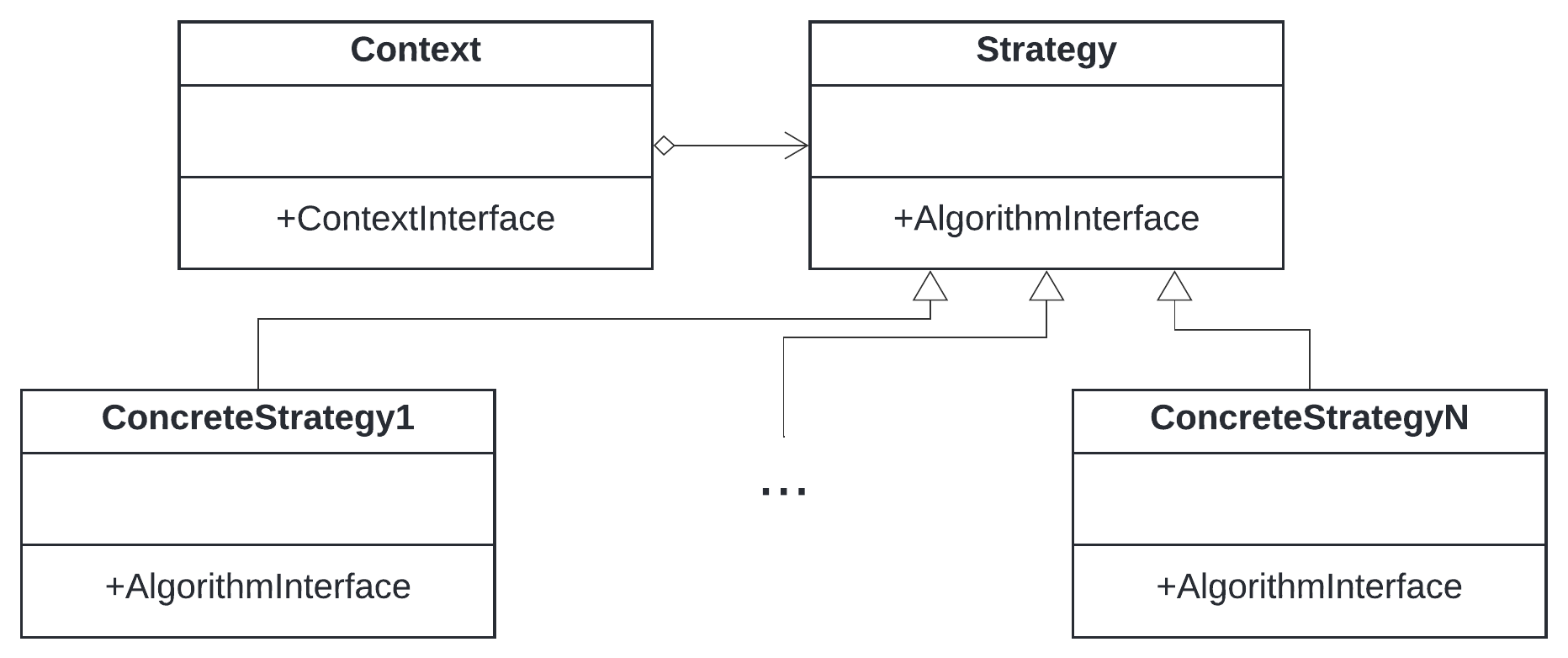}
    \caption{UML diagram of the \textit{Strategy} pattern.}
    \label{fig:strategypattern}
\end{figure}

The \textit{Strategy} pattern was first proposed in the world-famous book by Gamma et al.~\cite{designpatternsgamma} on design patterns for object-oriented programming. 
The \textit{Strategy} design pattern is a behavioural pattern exploiting a family of algorithms that offer a same functionality, enabling the dynamic selection of those algorithms at runtime, independently from clients and based on user preferences, software configuration, or  current system states. 
Strategies are each implemented in a separate class and offer the \textsf{\small Strategy} interface, exploited then by a \textsf{\small Context} class. The \textsf{\small Context} provides an interface between the application's client and the different available algorithms, by selecting the most appropriate one to serve incoming requests.
Particularly, as sketched in the UML diagram of Fig.~\ref{fig:strategypattern}:

\begin{itemize}
    \item the \textsf{\small Context} links to a \textsf{\small Strategy} object and delegates tasks (e.g., \rev{HTTP} requests, method invocations) to it, acting as an interface (i.e. the \textsf{\small ContextInterface}) between the client and the \textsf{\small Strategy} object,
    \item the \textsf{\small Strategy}, in turn, is an interface (i.e. the \textsf{\small AlgorithmInterface}) to be implemented by a set of concrete strategies, which can be useed interchangeably within the \textsf{\small Context}, depending on clients' needs, and
    \item the \textsf{\small ConcreteStrategy1}, $\dots$, \textsf{\small ConcreteStrategyN}  implement specific versions of an algorithm, adhering to the interface specified by the \textsf{\small Strategy}.
\end{itemize}

Such a pattern favours software flexibility, modularity, and reuse. It is then up to the clients to express their preferences and to select the best strategy to fulfil their needs at runtime.

We employ the \textit{Strategy} pattern to implement carbon-aware interactive \rev{software} services. We assume \rev{service} functionalities -- accessible through a REST API -- can be provided in a set of different flavours, each associated to a different result quality (e.g., average error), execution time, and, therefore, energy consumption. In general, the lower the execution time, the lower the result quality and energy consumption.

\begin{figure}
    \centering
    \includegraphics[width=0.49\textwidth]{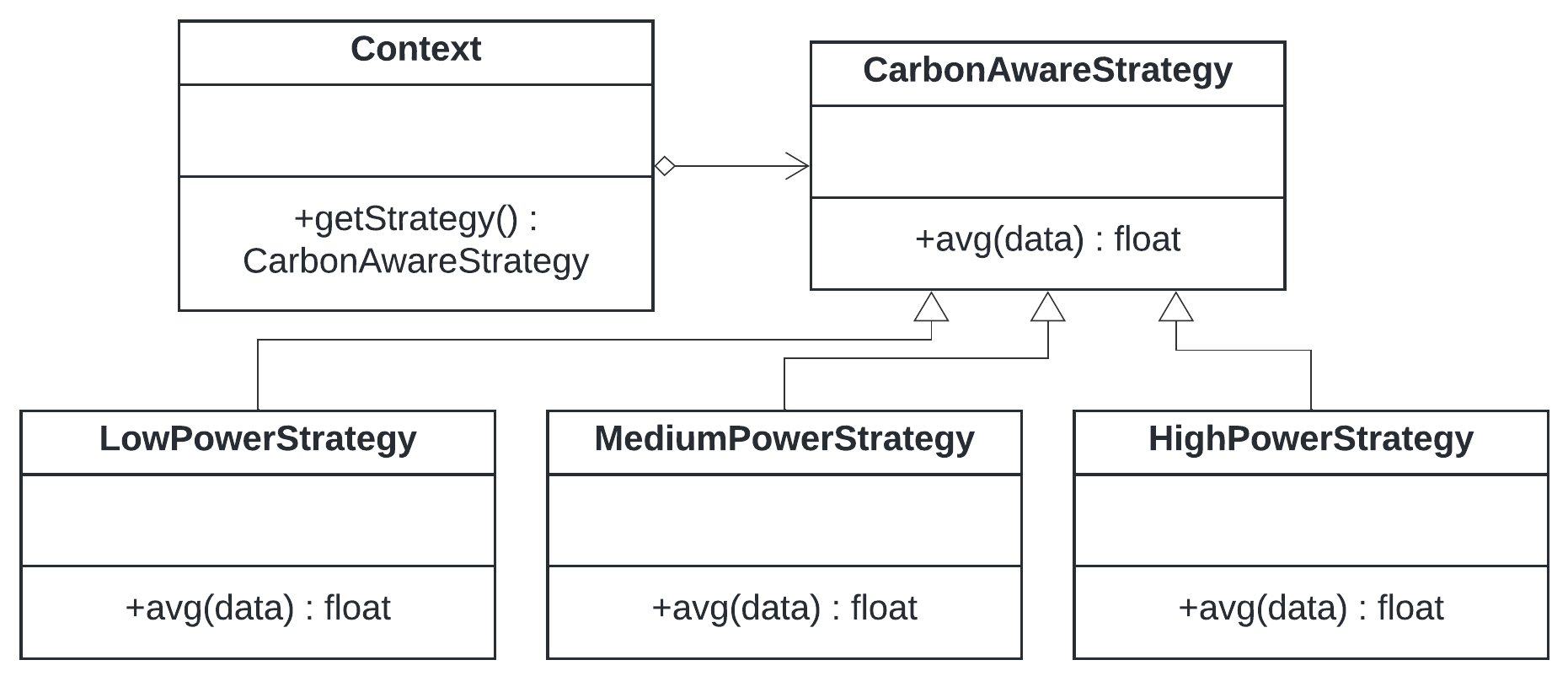}
    \caption{UML diagram of \carbostat case study.}
    \label{fig:spexample}
\end{figure}

Fig.~\ref{fig:spexample} and~\ref{fig:strategies} show an example of Strategy pattern within \carbostat. Consider a RESTful service featuring a single operation \texttt{/avg} that 
returns the average value of a given list of integers.
The \rev{service} is associated with a \textsf{\small Context} object that is in charge of selecting a suitable strategy to compute the average, by relying on the three concrete implementations available for the abstract interface \textsf{\small CarbonAwareStrategy}, viz. \textsf{\small LowPowerStrategy}, \textsf{\small MediumPowerStrategy}, \textsf{\small HighPowerStrategy}. 

Fig.~\ref{fig:interface} lists the Python code for the abstract class \textsf{\small CarbonAwareStrategy}. Such interface only offers the \texttt{avg(data)} method, which supposedly inputs a list of \texttt{data} and returns its average, and can be implemented into different classes, i.e. concrete strategies. By running controlled experiments, the application developers can associate each concrete strategy with its average execution time and average error on output results over the deployment servers\footnote{This can be achieved, for instance, by using the code provided at \url{https://github.com/di-unipi-socc/carbonstat/tree/main/data/time_error}}. \rev{The strategies we consider are defined in terms of the function \texttt{average(2)} listed in Fig.~\ref{fig:interface}. Such function inputs a list \texttt{data} of integers and a \texttt{step} value, which allows skipping some integers when accumulating values to compute the average of \texttt{data}.}

\rev{Overall, our service implements:}

\begin{itemize}
    \item a \textit{low-power} strategy that computes a statistical average over one fourth of the values in the list, by picking one every four items in the list (Fig.~\ref{fig:lowpower}), with associated average execution time of 35.3 ms and an average error on the results of 13.4\%,
    \item a \textit{medium-power} strategy that also computes a statistical average over half of the values in the list, by picking one every two items in the list (Fig.~\ref{fig:medpower}), with associated average execution time of 66.3 ms and an average error on the results of 4.5\%, and
    \item a \textit{high-power} strategy which computes the exact average over all values in the list (Fig.~\ref{fig:highpower}), with associated average execution time of 100.2 ms and no error on output results.
\end{itemize}

\begin{figure}[!h]
\centering
\begin{subfigure}{.45\textwidth}
  \centering
\begin{lstlisting}[language=Python,numbers=none]
class CarbonAwareStrategy(ABC):
    @abstractmethod
    def avg(data) -> float:
        pass

    def avg(data, step=1):
        sum = 0
        count = 0
        size = len(data)
        for i in range(0,size,step):
            count += 1
            sum += data[i]
        return round(sum/count)
\end{lstlisting}\caption{\textit{abstract carbon-aware strategy}}\label{fig:interface}\medskip
\end{subfigure}%
\\
\begin{subfigure}{.45\textwidth}
  \centering
\begin{lstlisting}[language=Python,numbers=none]
class LowPowerStrategy(CarbonAwareStrategy):
    def avg(data) -> float:
        return avg(data, step=4)
        
\end{lstlisting}\caption{\textit{low-power concrete strategy}}\label{fig:lowpower}\medskip
\end{subfigure}
\\
\begin{subfigure}{.45\textwidth}
  \centering
\begin{lstlisting}[language=Python,numbers=none]
class MediumPowerStrategy(CarbonAwareStrategy):
    def avg(data) -> float:
        return avg(data, step=2)
\end{lstlisting}\caption{\textit{medium-power concrete strategy}}\label{fig:medpower}\medskip
\end{subfigure}
\\
\begin{subfigure}{.45\textwidth}
  \centering
\begin{lstlisting}[language=Python,numbers=none]
class HighPowerStrategy(CarbonAwareStrategy):   
    def avg(data) -> float:
        return avg(data)
\end{lstlisting}\caption{\textit{high-power concrete strategy}}\label{fig:highpower}\medskip
\end{subfigure}%
\\
\caption{Strategies of \carbostat case study.}
\label{fig:strategies}
\end{figure}

The above strategies are offered to end-users as an interactive service to compute the average of a list of integers. 
Fig. \ref{fig:context} shows the \texttt{\small Context} class, fed with a strategy \texttt{\small assignment} array (line 4) by the \texttt{\small \_\_init\_\_()} constructor, which loads a pre-computed assignment of strategies to time slots within a certain period of time (lines 5--7). 
The  \texttt{\small Context} then exposes a \texttt{\small getStrategy()} method (lines 9--13) that either returns the strategy set for the current time (line 12--13), or a specific strategy, \texttt{\small force\_strategy}, forced by the caller (i.e. lines 10--11), e.g., for those applications that do not tolerate errors on the computed results (e.g., life-critical applications).

To conclude our case study on how to design and implement a carbon-aware service,  Fig.~\ref{fig:caservice} lists the Flask code for our simple application. It exposes a REST endpoint featuring the average function implemented by the three strategies and invoked through the \texttt{\small Context} object (lines 18--27). Every incoming request is handled by the \texttt{\small avg()} function that first retrieves the strategy to be used from the \texttt{\small Context} instance (line 21). Note that a non-mandatory parameter \texttt{\small force\_strategy} can be associated to the request so as to force the execution of a specific concrete strategy (line 20). Last, the selected strategy is used to compute the average on the input data (line 23) before returning it to the end-user (lines 25--27).

In the next section, based on forecast carbon intensity of the energy mix at use and forecast request rates for our service, we will discuss how strategies can be mapped and used in different time slots across the day so as to keep the average error of output results above a set thresholds and, subsequently, to minimise carbon emissions.

\begin{figure}[!h]
\begin{lstlisting}[language=Python,numbers=right,numberblanklines=true,firstnumber=1]
class Context:
    
    def __init__(self):
        self.assignment = {}
        assignment = open(environ["ASSIGNMENT"])
        self.assignment = load(assignment)
        assignment.close()
    
    def getStrategy(self,force_strategy) -> CarbonAwareStrategy:
        if force_strategy is not None:
            return CarbonAwareStrategies[force_strategy].value
        now = datetime.now()
        return self.assignment[now.hour + 0.5*floor(now.minute/30)]
\end{lstlisting}
\caption{Context implementation.}
\label{fig:context}
\end{figure}

\vspace{-0.75cm}
\begin{figure}[!h]
\begin{lstlisting}[language=Python,numbers=right,numberblanklines=true,firstnumber=14]
app = Flask(__name__)

app.context = Context()

@app.route("/avg")
def avg():
    force_strategy = request.args.get("force")
    strategy = app.context.getCarbonAwareStrategy(force_strategy)

    average = strategy.avg(app.data)

    result = {}
    result["value"] = average
    return jsonify(result)

\end{lstlisting}
\caption{Carbon-aware service implementation.}
\label{fig:caservice}
\end{figure}

\vspace{-0.7cm}
\section{Setting Strategies, with \carbostat}
\label{sec:carbomilp}\noindent
\noindent
In this section, given an interactive service implemented as described in Sect.~\ref{sec:design}, we illustrate how our \carbostat prototype can be used to configure set-points of the service \texttt{\small Context} so to select the best strategy according to forecast requests and carbon intensity of the energy mix that powers up the host servers. In detail, we first present the formulation of our problem as a bilevel optimisation (Sect.~\ref{sec:bilevel}), then illustrate it through a running example (Sect.~\ref{sec:runningexample}).

In what follows, we assume that predictions on both service demand and carbon intensity are available within a time horizon (e.g. 24--48 hours). Many accurate techniques exist to predict such values, like \cite{qiu2019dynamic,nishimatsu2006service} for service demand and \cite{maji2023multi,leerbeck2020short} for carbon intensity.

\subsection{Problem Formulation}
\label{sec:bilevel}
\noindent
Consider a time interval represented by a sequence of $t$ discrete time slots, for which a forecast of carbon emissions $c_i$ and user requests $r_i$ is available (with $i=1,\dots,t$). 
Suppose that the target service is available with $s$ implemented strategies, each with its own average error $e_j$ and execution time $d_j$ (with  $j=1, \dots, s$).
We aim at assigning a strategy to each time slot to minimise the overall carbon emissions and keep the average error of the results returned by the target service within a tolerated error threshold $\varepsilon$. Among the assignments that minimise carbon emissions, we then aim at minimising the average error of returned results.

By exploiting the notation of \Cref{tab:problem-notation}, the above is formalised by the following bilevel optimisation~\cite{Sinha18_BilevelOptimizationReview} problem, defined over the binary decision variable $x_{ij}$ \rev{that is set to $1$ to indicate that strategy $j$ is assigned to time slot $i$, to $0$ otherwise:}

\begin{small}
\begin{equation}\label{eq:minerror}
\min_{x \in X} \sum_{i=1}^{t}\sum_{j=1}^{s} x_{ij}r_ie_j
\end{equation}
\end{small}%
subject to
\begin{small}
\begin{equation}\label{eq:mincarbon}
X = \arg\min_{x \in Y} \Big\{\sum_{i=1}^{t}\sum_{j=1}^{s} x_{ij}c_i r_i d_j \Big\}
\end{equation}
\end{small}%
where
\begin{small}
\begin{equation}
\label{eq:avg-error}
Y = \Big\{ x \in V \mid \sum_{i=1}^{t}\sum_{j=1}^{s} x_{ij}r_ie_j \leq \varepsilon \sum_{i=1}^{t}\sum_{j=1}^{s} x_{ij}r_i \Big\}
\end{equation}
\begin{equation}
\label{eq:assignment-matrix}
V = \Big\{ x  \in \{0,1\}^{t \times s} \, \mid \, \sum_{j=1}^{s} x_{ij} = 1, \ i=1,\dots,t \Big\}
\end{equation}
\end{small}

\begin{table}[]
    \centering
    \caption{Notation recap.}
    \label{tab:problem-notation}
    \begin{tabular}{|cl|}
        \hline
        \textbf{Symbol} & \textbf{Meaning} \\
        \hline
        $\varepsilon$ & tolerated average error \\
        $x$ & assignment matrix \\
        $x_{ij}$ & assignment of strategy $j$ to time slot $i$\\
        $t$ & total number of time slots \\
        $c_i$ & estimated carbon emissions for time slot $i$ \\
        $r_i$ & user requests received at time slot $i$ \\
        $s$ & total number of available strategies \\
        $d_j$ & average duration for strategy $j$ \\
        $e_j$ & average error for strategy $j$ \\
        \hline
    \end{tabular}
\end{table}

\noindent
Solutions to the above problem are within the domain of binary $t \times s$ matrices denoting assignments of strategies to time slots, with each time slot obviously assigned with a single strategy (\cref{eq:assignment-matrix}). 
We further restrict assignments to those maintaining the average error within $\varepsilon$ (\cref{eq:avg-error}).
Then, the lower-level optimisation task minimises the overall carbon emission (\cref{eq:mincarbon}), and the upper-level optimisation task minimises the average error of the results returned by the target service (\cref{eq:minerror}).

The above formulation makes it clear that the upper-level optimisation task strictly depends on determining the optimal Pareto front for the lower-level optimisation task. This naturally induces a priority between tasks, which requires to first solve the lower-level and, then, the upper-level one. It is worth noting that this type of problems are inherently NP-hard~\cite{dempe2020bilevel}.

\subsection{\carbostat at work}
\label{sec:runningexample}

\noindent
The considered problem was encoded in an open-source Python prototype\footnote{Available at: \url{https://github.com/di-unipi-socc/carbonstat/blob/main/carbonstat/carbonstat.py}}, \carbostat, by relying on Google's optimisation suite OR-tools\footnote{\textit{Google OR-tools}, \url{https://developers.google.com/optimization}}. 

Our prototype first determines all possible solution $x$ optimising the inner objective set by Eq.~\ref{eq:mincarbon} and meeting Eq.~\ref{eq:avg-error} and Eq.~\ref{eq:assignment-matrix}. Among these, it then determines one solution that also optimises the outer objective of Eq.~\ref{eq:minerror}. 
We now show the solutions it finds over a small motivating example considering six time slots only. The prediction values for those time slots are as follows:

 $$r=\icol{350\\500\\1000\\750\\400\\100}\ \ \ c = \icol{260\\350\\220\\530\\610\\1100}$$

\noindent where, for instance,  at time slot $i=1$ there is a forecast of 350 incoming requests and of a carbon intensity of 260 g$\text{CO}_2\text{-eq/kWh}$, while at  $i=6$  there is a forecast of 100 incoming requests and of a carbon intensity of 1100 g$\text{CO}_2\text{-eq/kWh}$. 

We assume the strategies that can be selected are those of the case study of Sect.~\ref{sec:design}, that is they are denoted by the following average execution times and error on output results:

 $$d=\icol{35.3\\66.3\\100.2}\ \ \ e = \icol{13.4\\4.5\\0}$$

 By using \carbostat over the above input data, we can solve the bilevel optimisation problem of Sect.~\ref{sec:bilevel} for different values of the maximum tolerated error $\varepsilon$.

 Running our prototype after setting the maximum tolerated error $\varepsilon=0\%$, naturally returns the following optimal assignment, which always exploits the \enquote{zero-error} strategy, i.e., the \textit{high-power} strategy:

 $$x=\begin{pmatrix}
0 & 0 &  1 \\
0 & 0 &  1 \\
0 & 0 &  1 \\
0 & 0 &  1 \\
0 & 0 &  1 \\
0 & 0 &  1  \\
\end{pmatrix}$$

Considering running our service on a 50W server, this configuration emits 1.72 gCO$_2$-eq over the considered time slots. This first solution incurs no error but, on the other hand, gives no control over the produced carbon emissions.
Similarly, setting $\varepsilon=15\%$ we obtain the following assignment, which always runs the \textit{low-power} strategy -- as 15\% is above the 13.4\% average error of the \textit{low-power} strategy:

$$x=\begin{pmatrix}
1 & 0 &  0 \\
1 & 0 &  0 \\
1 & 0 &  0 \\
1 & 0 &  0 \\
1 & 0 &  0 \\
1 & 0 &  0  \\
\end{pmatrix}$$

Considering again running our service on a 50W server, this configuration emits 0.6 gCO$_2$-eq over the considered time slots, i.e., 65.1\% less than the \textit{high-power} configuration. This \textit{low-power} configuration, naturally, incurs the highest average error on output results of $13.4\%$. In contrast with the previous case, this configuration minimises emissions but gives no control over the error on output results.

Last, setting $\varepsilon=5\%$, we get the following optimal assignment, with an average error on output results of $4.98\%$ and minimising overall carbon emissions:

 $$x=\begin{pmatrix}
0 & 0 &  1 \\
0 & 1 &  0 \\
0 & 0 &  1 \\
1 & 0 &  0 \\
0 & 1 &  0 \\
1 & 0 &  0  \\
\end{pmatrix}$$

Such assignment exploits the \textit{high-power} strategy at time slots $1$ and $3$, the \textit{medium-power} strategy at time slots $2$ and $5$, and the \textit{low-power} strategy at time slot $4$ and $6$. Notably, \carbostat selects the best strategy to handle situations in which a lower carbon intensity allows to handle large amounts of requests at \textit{high-power} (e.g. time slot $i=3$) or, vice versa, a higher carbon intensity in spite of few requests is better handled at \textit{low-power} (e.g. time slot $i=6$). 
Considering again running our service on a 50W server, this configuration emits 1.07 gCO$_2$-eq, i.e. 37.8\% less than the \textit{high-power} configuration and 43.9\% more than the \textit{low-power}. To conclude this example, note that \carbostat allows to control the average quality of output results while minimising carbon emissions as much as possible.

\section{Evaluation}
\label{sec:experiments}\noindent
In this section, we illustrate and discuss the results of a controlled experiment\footnote{All experimental code and data are publicly available on GitHub at \url{https://github.com/di-unipi-socc/carbostat/tree/main/data/experiment}. Experimental results (included average execution times and error for the considered strategies) were obtained by running the suitably configured average service over a Ubuntu 20.04 LTS server, equipped with 4 CPUs and 32 GB of RAM. Experiments were repeated two times to mitigate experimental errors.} executed by relying on the service described in Sect.~\ref{sec:design}, configured by running \carbostat over \rev{realistic carbon intensity data and request load patterns.} 

\smallskip 
\noindent
\textbf{Setup}.
\noindent Our experiments aimed at comparing the amount of carbon emissions and the average error of running the carbon-aware average \rev{software} service (\Cref{sec:design}), with each time slot's strategy selected according to seven different configurations, against a standard version featuring only the \textit{high-power} strategy. Namely, we considered the following configurations:
\begin{itemize}
    \item \alwayslow and \alwaysmedium, which always run the \textit{low power} and \textit{medium power} strategy of the service, respectively, independently of the carbon intensity or user requests of each time slot, 
    \item \naive, which runs the \textit{high power} strategy of the service in a time slot if the carbon intensity \rev{is low} and user requests are below 1/3 of the maximum expected number of user requests, the \textit{medium power} strategy if the carbon intensity is at most moderate and user requests are below 2/3 of the maximum expected number of user requests, and the \textit{low power} strategy in any other case.\footnote{The thresholds for \enquote{low} or \enquote{moderate} carbon intensity were aligned to those of the Carbon Intensity API \cite{NationalGridESO2024_CarbonIntensityAPI}.}
    \item \carbostat($\varepsilon=x$), with $x \in \{1\%,2\%,4\%,8\%\}$, which assigns strategies to time slots by running \carbostat with the tolerated error threshold set to $x$.
\end{itemize}

\begin{figure}
    \centering
    \includegraphics[width=.49\columnwidth]{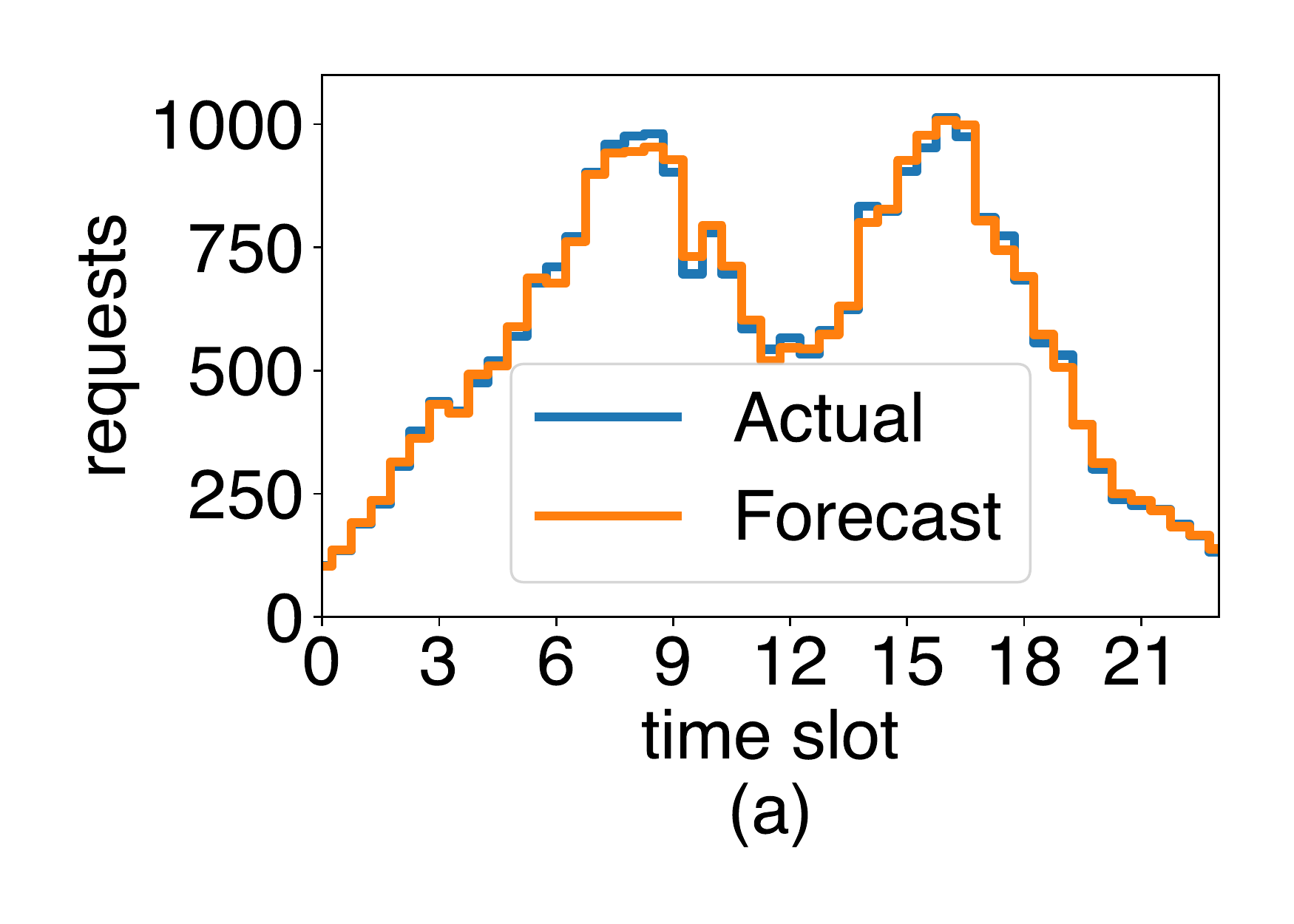}
    \includegraphics[width=.49\columnwidth]{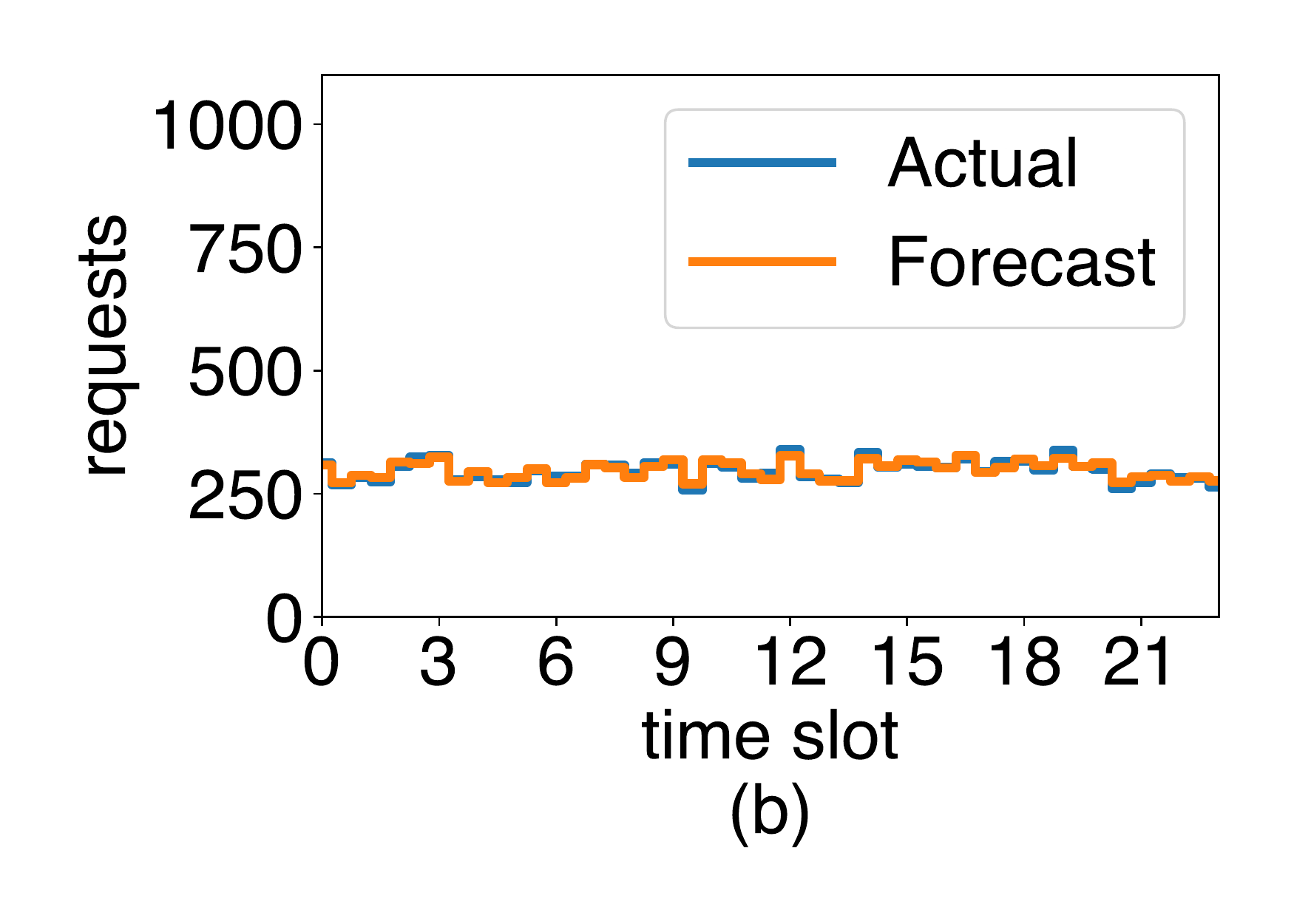}
    \\
    \includegraphics[width=.49\columnwidth]{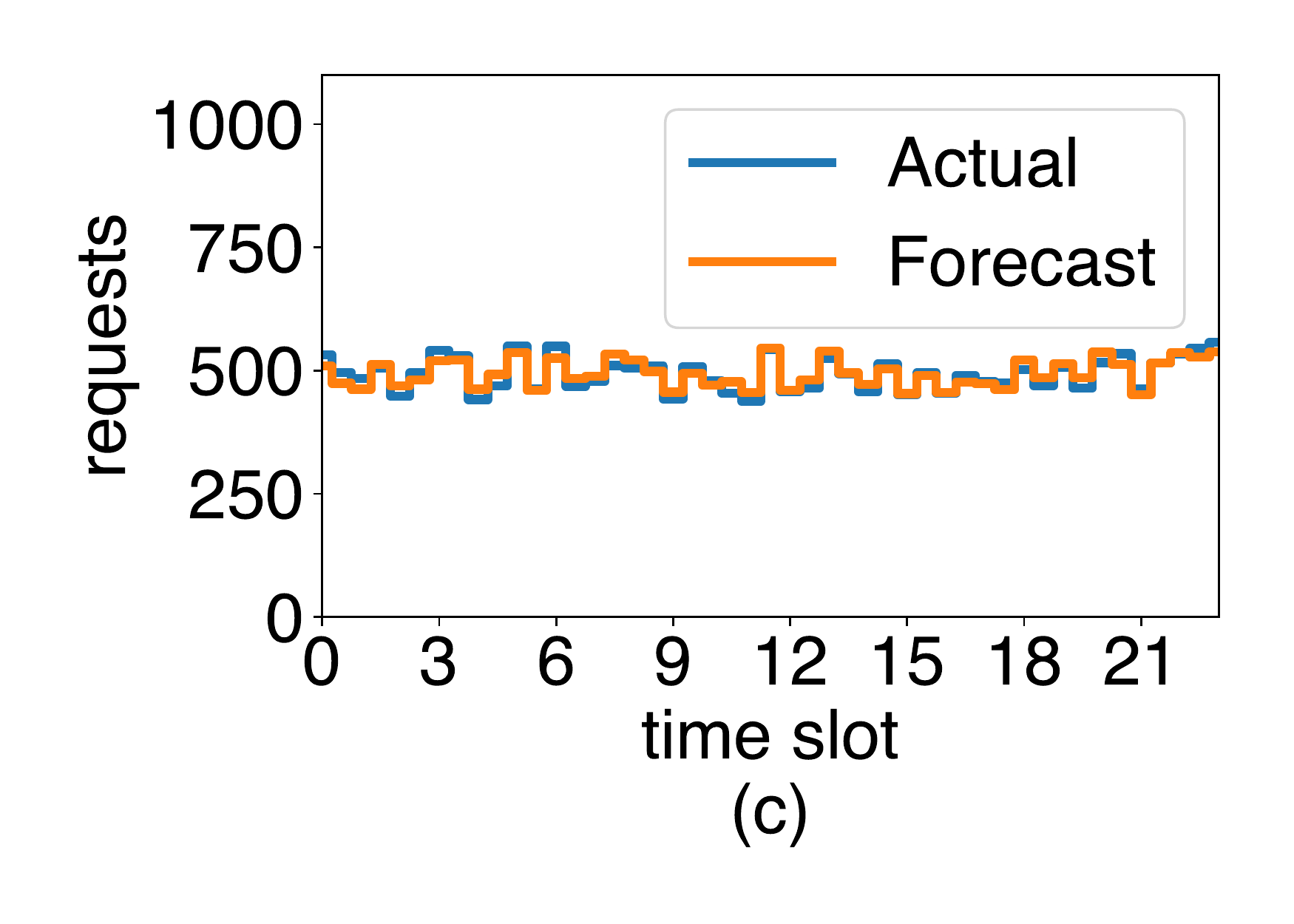}
    \includegraphics[width=.49\columnwidth]{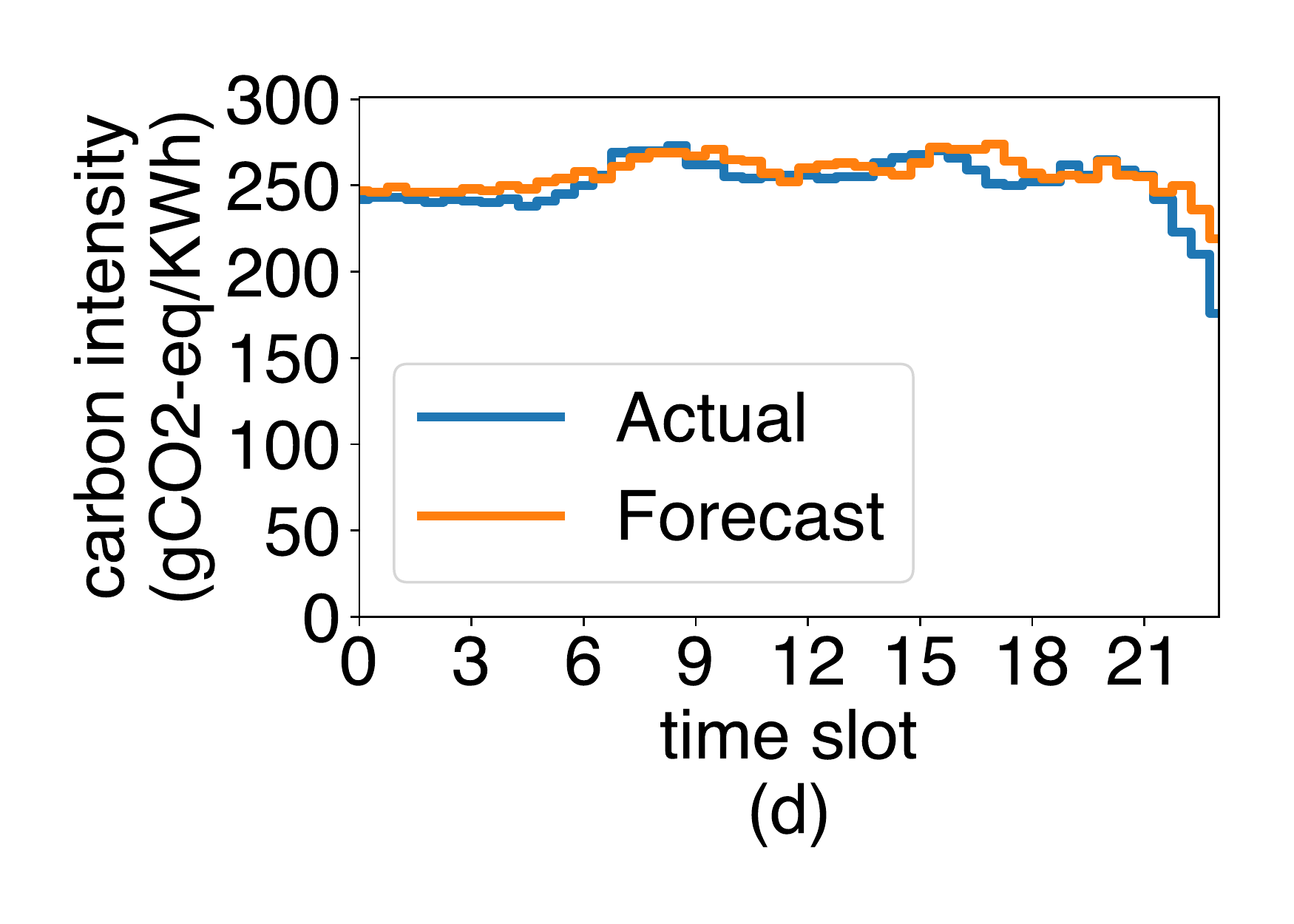}
    \caption{Example of evolution of the values of (a-c) user requests and (d) carbon intensity in one of the considered days.}
    \label{fig:experiment-input-values}
\end{figure}

We first measured the \textit{average error} of the output results and \textit{service time} for each strategy of the approximated average service, with the service time being the time between the service receiving a request and sending its corresponding reply.
We then considered 12 different days of the year 2023, each taken from a different month so to account for a meaningful variety of possible carbon intensities. The values of forecast and actual carbon intensity for each 30-minutes time slot of each considered day were obtained via the Carbon Intensity API \cite{NationalGridESO2024_CarbonIntensityAPI}. For instance, \Cref{fig:experiment-input-values}(d) plots forecast and actual carbon intensity values for a day in January 2023. 

Instead, the values of forecast and actual user requests per each day's time slot were randomly generated by means of three different probability distributions following three lifelike requests' profiles. Namely, we exploited:

\begin{itemize}
    \item a profile with peaks around 1000 requests occurring at early in the morning and late in the afternoon, see \eg \Cref{fig:experiment-input-values}(a), 
    \item a profile with stable load around 300 requests, see \eg \Cref{fig:experiment-input-values}(b), and
    \item a profile with stable load around 500 requests, see \eg \Cref{fig:experiment-input-values}(c).
\end{itemize}

\noindent\rev{Actual profiles were obtained by perturbing the amounts of requests specified above by a random variance of $\pm 10\%$}.

For each considered day, we passed the forecast values of carbon intensity and user requests, and the strategies' average service time and error, as input to \carbostat to obtain the assignments of \carbostat ($\varepsilon=x$), $x \in \{1\%,2\%,4\%,8\%\}$.
We used the actual values of carbon intensity from the Carbon Intensity API and user requests obtained by varying forecast requests by a random amount between 0 and $\pm 5\%$ to simulate the execution of the approximated average service on each time slot of each considered day as if strategies were assigned by each of the seven considered policies. A time slot was simulated by invoking the deployed instance of the approximated average service for a number of times equal to the time slot's actual user requests, and by assuming that they were served by consuming energy produced with the time slot's actual carbon intensity. 

We then measured the average error and overall carbon emissions as the average and sum of those measured on each time slot, respectively.
The error was computed as a percentage denoting the distance from the returned average and the actual one, with the latter obtained by initially forcing a request to be served with the \textit{high power} strategy. The carbon emissions for serving each request, instead, were estimated based on the service time for such request. 
More precisely, the carbon emissions for each request were obtained by multiplying the corresponding service time by the time slot's carbon intensity and by the server's power consumption (i.e., 50W). 

\begin{figure}
    \centering
    \includegraphics[width=.98\columnwidth]{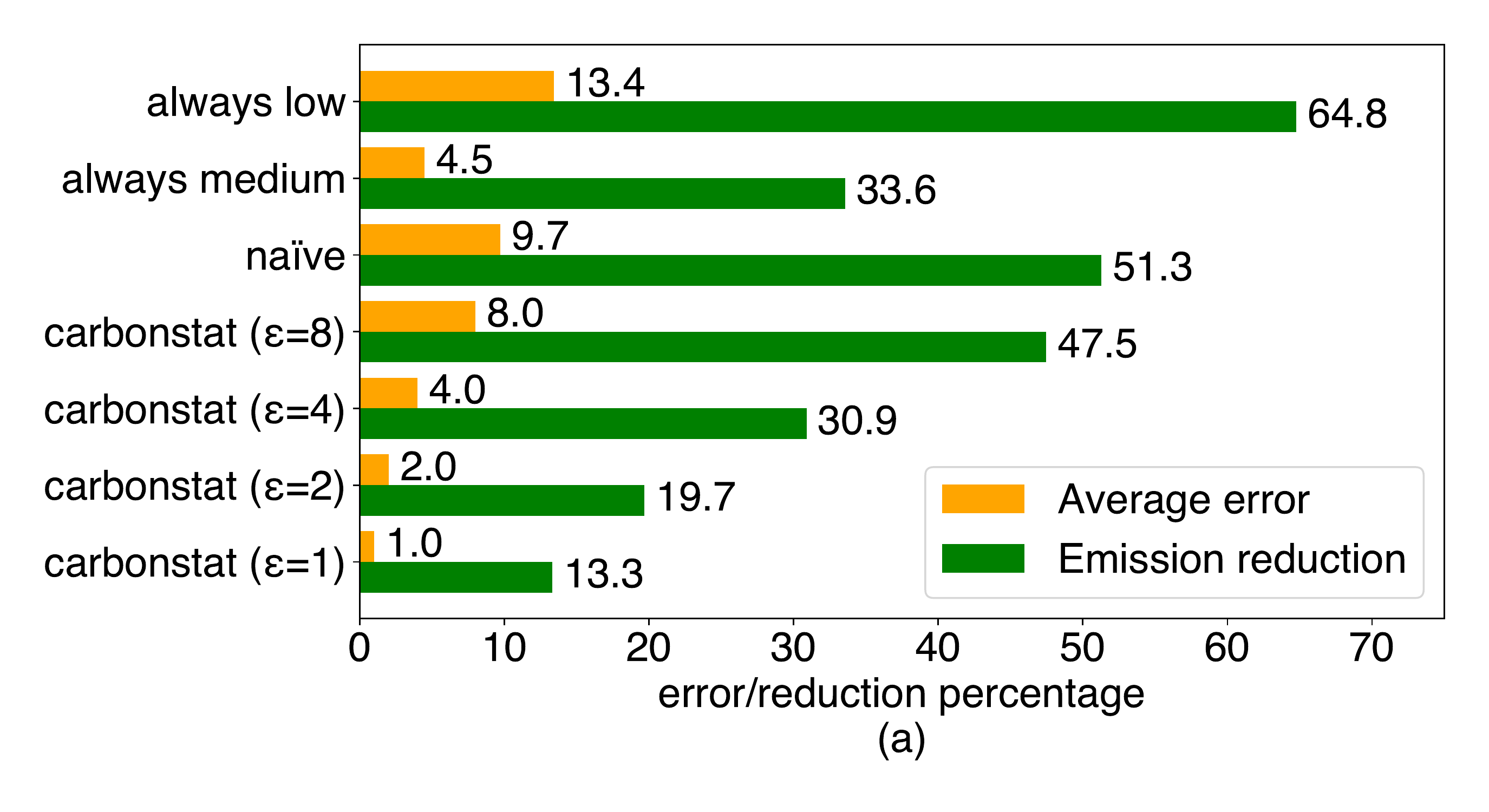}
    \\
    \includegraphics[width=.98\columnwidth]{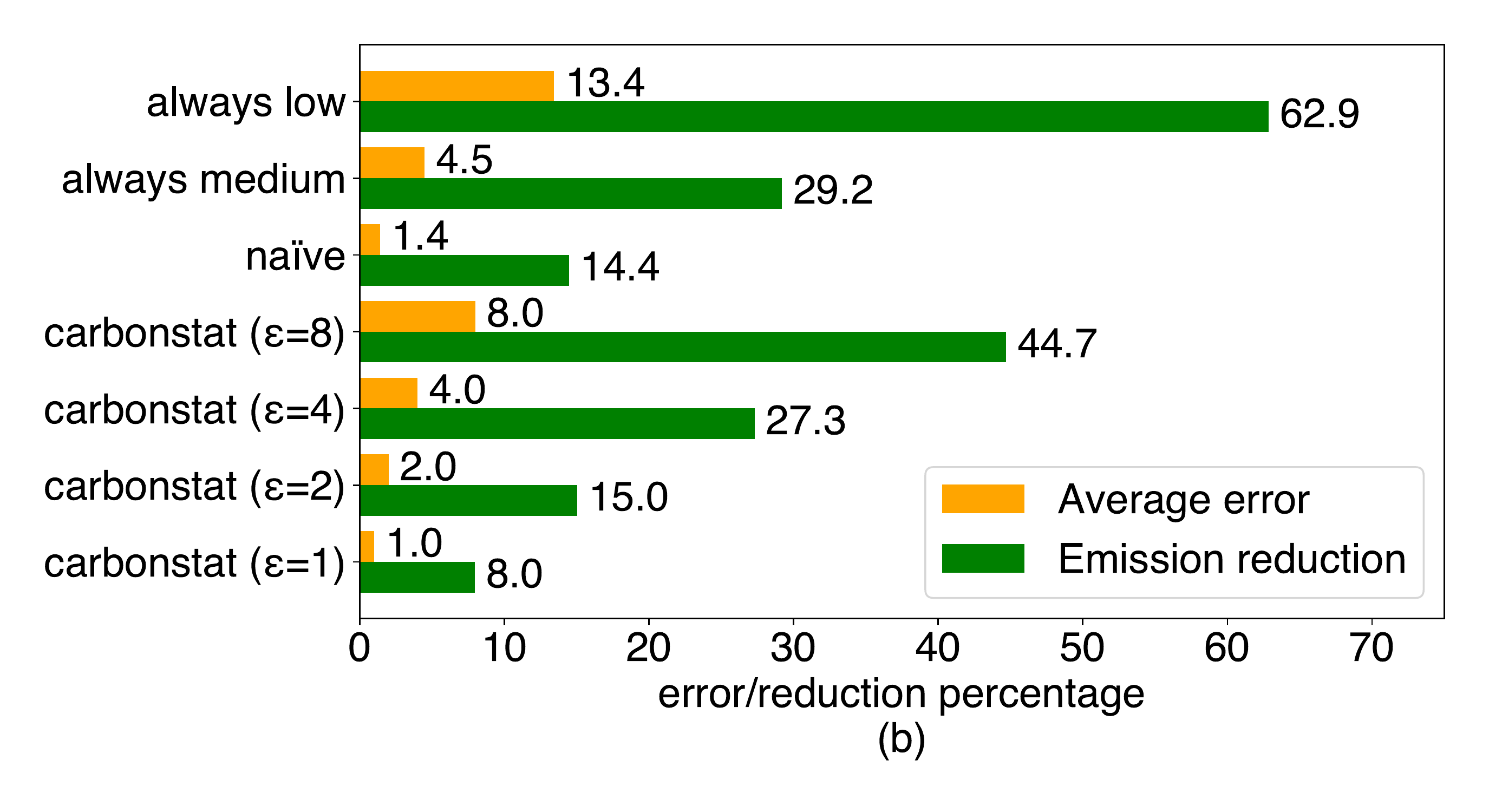}
    \\
    \includegraphics[width=.98\columnwidth]{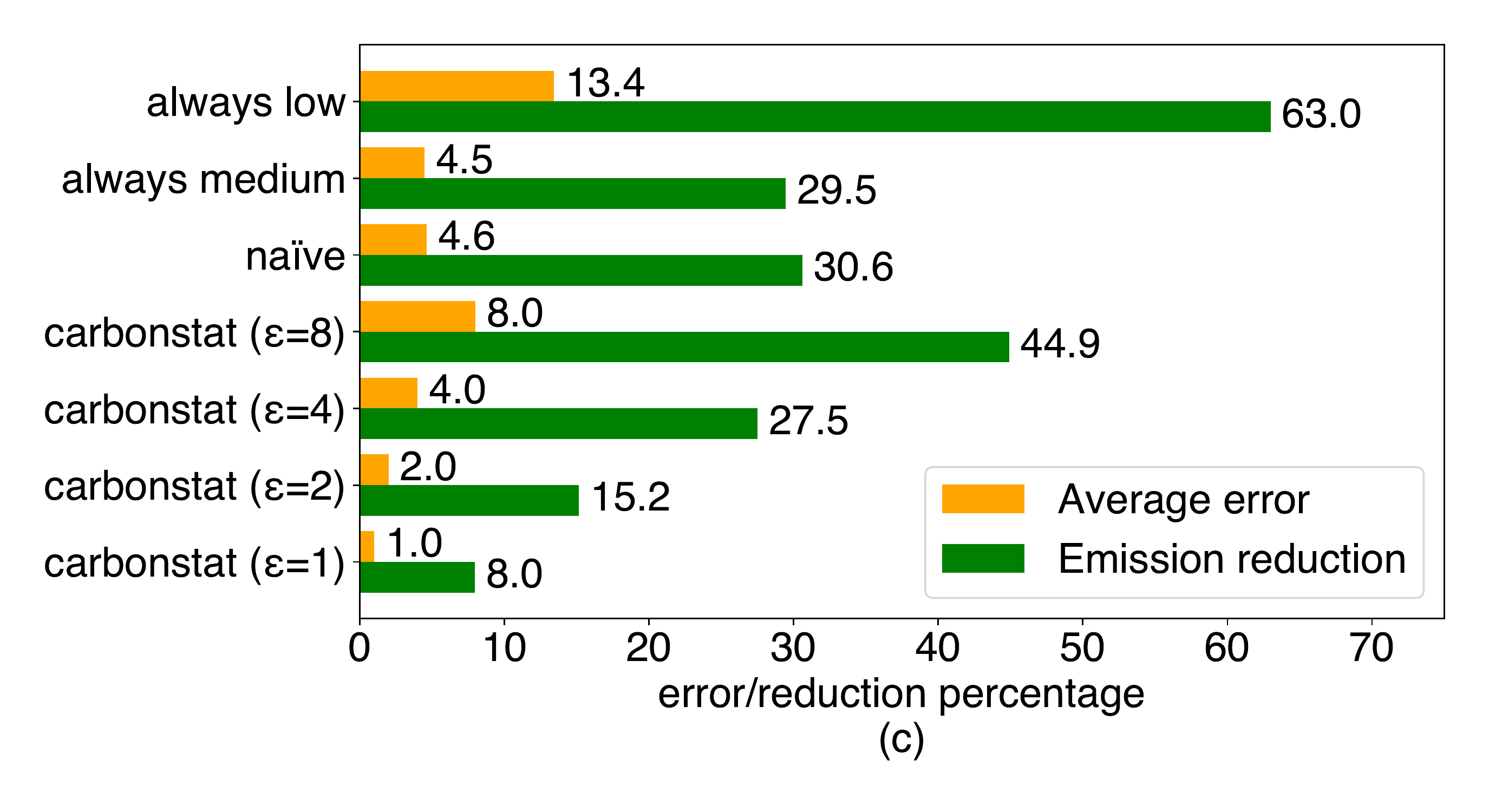}
    \caption{Experimental results.}
    \label{fig:experiment-results}
\end{figure}

\smallskip 
\noindent
\textbf{Results}.
\Cref{fig:experiment-results} illustrates the average results obtained in our experiments for each given configuration, in terms of average error and carbon emissions reduction in comparison to the service only featuring the \textit{high-power} functionality. Particularly, 
\Cref{fig:experiment-results}(a), (b), and (c) concerns the experiments corresponding to the request profiles sketched in \Cref{fig:experiment-input-values}(a), (b) and (c), respectively.

In all experiments, as expected, the \alwayslow and the \alwaysmedium strategies show an average error on output results equal to that of the corresponding strategies, viz., $13.4\%$ and $4.5\%$, respectively. Similarly, across all experiments, the configurations obtained via \carbostat guarantee that the average error on output results settles exactly around the set maximum error threshold $\varepsilon \in \{1\%, 2\%, 4\%, 8\%\}$. 

It is worth analysing how the average desired average error threshold is not exceeded by the \carbostat instances across the three experiments, i.e. guaranteeing the desired quality for output results. \rev{With $\varepsilon=1\%$ as the maximum average tolerated error on output results, \carbostat($\varepsilon$=1) is the sole configuration} that can guarantee such average error, also reducing carbon emissions with respect to the \textit{high-power} baseline. Indeed, independently of the considered request profile, \carbostat($\varepsilon$=1) guarantees an average error of $1\%$ and reduces emissions by 13.3\% in for the peaky request profile (a) and by $8\%$ for the stable request profiles (b) and (c).

\rev{When considering $\varepsilon=2\%$ as the maximum average tolerated error on output results, \carbostat($\varepsilon$=2) is the only configuration} capable of satisfying such requirement both in the case of peaky load (a) and stable load of 500 requests (c), also achieving a carbon reduction of 19.7\% and 15.2\%, respectively. In the case of stable load profile of 300 requests (b), both the \naive and the \carbostat($\varepsilon$=2) configuration manage to meet the error threshold. However, \carbostat($\varepsilon$=2) achieves a \rev{slightly} higher reduction of carbon emissions ($15\%$) in comparison with the \naive approach (14.4\%). \rev{Naturally, also \carbostat($\varepsilon$=1) meets the requirement on the average error of $2\%$ for all request profiles, at the price, however, of a lower reduction of carbon emissions in all considered scenario. }

If $\varepsilon=4\%$, \carbostat($\varepsilon$=4) still is the only approach to meet the maximum error threshold for peaked requests (a) and stable requests (c), reducing carbon emissions by $30.9\%$ and $27.5\%$, respectively. In the case of stable requests (b), the \naive policy also does not exceed the maximum tolerated error but has a reduction in carbon emissions which is around half the one achieved by \carbostat($\varepsilon$=4) (14.4\% vs 27.3\%). \rev{As before, \carbostat($\varepsilon$=1) also meets the requirement on the average error of $4\%$, producing higher carbon emissions than \carbostat($\varepsilon$=4), in (a), (b), and (c), and \naive in (b). Similarly, \carbostat($\varepsilon$=2) satisfies the requirement set on the average error in all cases, with higher emissions than \carbostat($\varepsilon$=4) for request profiles (a), (b) and (c), and with higher error than \naive for request profile (b) at a similar reduction in the carbon emissions.}

Last, when considering  $\varepsilon=8\%$ as the maximum average tolerated error on output results, the \carbostat configuration still works fine with all requests profiles with a reduction of emissions of $47.5\%$ for peaky requests (a), and of $44.7\%$ for stable request loads (b) and (c). The \alwaysmedium configuration works for the peaky load (a) but causes a more modes reduction in carbon emissions, settling around 33.6\%. 
The \alwaysmedium and \naive configurations also meet the 8\% set-point for request profiles (b) and (c), however they reduce carbon emissions less, when compared with \carbostat configuration. Particularly, \naive achieves a reduction of $14.4\%$ in case (b) and of 30.6\% in case (c), while \alwaysmedium reaches a reduction of $29.2\%$ in case (b) and $29.5\%$ in case (c). \rev{Again, the configuration obtained with \carbostat($\varepsilon$=1), ($\varepsilon$=2), and ($\varepsilon$=4) would support the requirement on the average error of at most $8\%$ incurring in lower reduction in carbon emissions w.r.t. \carbostat($\varepsilon$=8), i.e. without fully exploiting the potential reduction in carbon emissions that such tolerated error can enable.}

\smallskip 
\noindent
\textbf{Discussion}.
\noindent Our experimental assessment against varying carbon intensity of the available energy mix and of the incoming requests shows that our framework can: 
\begin{itemize}
    \item[(\textit{i})] guarantee average error rates within the set thresholds, both in case of constant and peaked requests' load profiles,
    \item[(\textit{ii})] dynamically adapt the behaviour of the configured service to minimise carbon emissions, while naïve static configuration might fail on this objective (as, e.g.,  in \Cref{fig:experiment-results}(b)), and
    \item[(\textit{iii})] consistently balance the performance of the service and its carbon footprint, achieving a reduction of carbon emissions between 8\% and 50\% in the considered scenarios.
\end{itemize}

Speaking of static policies, like \alwayslow, they can further minimise carbon emission in comparison to \carbostat, at the price of higher error rates. Analogously, naïve thresholds do not dynamically adapt to different service contexts, viz. energy mix and request rates.

\section{Related Work}
\label{sec:related}
\noindent 
Managing or adapting running software so to reduce its energy consumption and carbon emissions has been cyclically considered in the literature, with different goals and perspectives. Recently, much work has been carried out in the field of orchestrating application services in a sustainable manner. We refer the readers to our recent survey~\cite{GaglianeseGreenOrch2023} for more details about green application orchestration, while we focus here on techniques to implement software which itself implements an energy- and carbon-aware adaptive behaviour.

Bunse \& H{\"{o}}pfner~\cite{Bunse08_ResourceSubstitutionComponents} were among the first to focus on energy-awareness of communication within component-based software. They propose the development and usage of an \textit{energy management component} (EMC) to handle communication via resource substitution strategies (e.g., caching, replication or hoarding) to balance the trade-offs between energy, performance and hardware usage (\eg compute results vs use cached values vs invoke a remote service). The framework was later assessed on the dynamic adaptation of sorting algorithms, considering execution times~\cite{BunseHRM09Sorting}.  Despite relying on the \textit{Strategy} pattern, they require, differently from us, the deployment of the EMC, the dynamic (un)deployment of components and does not consider carbon emissions nor incoming requests to optimise software configuration.

Still only considering energy usage optimisation and Software-as-a-Service architectures, Alvares de Oliveira \& Ledoux~\cite{alvaresledoux2010} propose a framework targeting BPEL-based service compositions, based on an orchestration component to trigger reconfiguration events and make application deployment more sustainable in terms of energy footprint.  
Hasan et al.~\cite{DBLP:conf/cloudcom/HasanOLP16,DBLP:journals/tsusc/HasanOLP17} propose an autonomic self-adaptive auto-scaler for Software-as-a-Servive architectures, which accounts for response times, application availability and energy quality (green vs brown). Also Xu et al.~\cite{DBLP:journals/tsusc/XuTB21} study how to adapt Cloud application deployments based on information about the current energy mix and (interactive/batch) workload. Differently from our work, their focus is how to manage applications by reacting to monitored events through external controllers, instead of configuring them based on context predictions. \rev{Based on energy consumption and QoS, Kavanagh et al.~\cite{DBLP:journals/tsusc/KavanaghDEBG20} focus on runtime migration of high-performance computing tasks across heterogeneous accelerated infrastructure resources.} 


Focussing on the design of energy-aware programming languages, Canino \& Liu~\cite{CaninoLiu17} devise a proactive and adaptive framework based on mixed typechecking to embed energy information in written programs, enabling different \textit{modes} (i.e. data types and corresponding behaviours)  -- based on contextual information. Their main goal is to enable battery- and temperature-awareness at the compiler level for applications running on mobile devices. Similarly, Sampson et al.~\cite{DBLP:conf/pldi/SampsonDFGCG11} and Han et al.~\cite{DBLP:conf/ets/HanO13} follow the line of approximate computing by proposing type systems that support different data precision levels. This, however, requires rethinking CPU design so as to support both precise and approximate data types in order to reduce energy consumption of running programs~\cite{approximatecomputingMittal16b}. 

\rev{To the best of our knowledge, this work is among the first to propose a software engineering methodology and an optimisation scheme to code and tune carbon-aware interactive software services. 
Differently from previous efforts, based on carbon and workload predictions, our work (\textit{i}) minimises deployment carbon emissions and (\textit{ii}) guarantees that the average quality of output results stays above a set-point.}

\section{Conclusions}
\label{sec:conclusions}
\noindent 

This article proposed a framework to implement and configure adaptive carbon-aware interactive \rev{software} services.  The framework is made of a methodology to implement adaptive services based on established software engineering principles (i.e. the \textit{Strategy} pattern) and on a MILP approach to optimally configure those services. The MILP approach has been released an open-source tool named \carbostat and the overall framework has been assessed by controlled experiments run over a case study.

\carbostat inputs a specification of a service implemented as a set of strategies, each associated to its own average execution time and quality of output result (e.g. error). Based on those values and on a prediction of request rates and carbon intensity of the energy mix at use within a future time period, \carbostat can determine which strategy to adopt at different slots in such a period so to (\textit{i}) minimise carbon emissions and (\textit{ii}) keep the average error on output results below a set threshold. \rev{Interestingly, our framework can also be used to enhance existing applications with carbon-aware features so as to reduce their environmental footprint.}

Our experimental assessment of a \rev{software} service implemented and tuned with \carbostat shows the feasibility of our framework and its capability to adapt the service behaviour in response to \rev{lifelike} carbon emissions and request patterns, with reduction on carbon emissions between 8\% and 50\%, corresponding to guaranteed average error rates between $1\%$ and $8\%$. We have also shown that static configurations do not provide the same flexibility and might end up with different behaviours depending on the service execution context.

As future work, we plan to:
\begin{itemize}
    \item \rev{add continuous service re-configuration to adapt the service behaviour in case of sudden changes in the forecast of carbon emissions or incoming requests, as well as in response to unexpected events,}
    \item \rev{evaluate our approach over more complex services and in distributed Cloud-Edge settings, relying on real applications and involving end-users to include an assessment of their Quality of Experience, and}
    \item \rev{extend our methodology to accommodate other class of services that can tolerate delays (e.g. batch processing) also including the possibility of shifting service execution in time to further save on carbon emissions.}
\end{itemize}

\smallskip\noindent
{\small
\textsc{Acknowledgements} This work has been partly funded by projects:
\textit{Energy-aware management of software applications in Cloud-IoT ecosystems} (RIC2021\_PON\_A18) funded over ESF REACT-EU resources by the \textit{Italian Ministry of University and Research} through \textit{PON Ricerca e Innovazione 2014--20};
\textit{FREEDA} (CUP: I53D23003550006), funded by the frameworks PRIN (MUR, Italy) and Next Generation EU;
\textit{OSMWARE} (UNIPI\_PRA 2022 64), funded by the University of Pisa, Italy. Computational resources provided by \textsf{\small computing@unipi}, a computing service provided by the University of Pisa. 
\rev{The Authors wish to thank Giandomenico Mastroeni for his  feedback on the mathematical model presented in this work.}}



\bibliographystyle{IEEEtran}
\bibliography{src/biblio.bib}

\end{document}